\documentclass[12pt]{iopart}  
\usepackage[]{graphicx}
\usepackage{graphicx}
\usepackage{dcolumn}
\usepackage{bm}

\begin{document}
\title{Transient current in a quantum dot asymmetrically 
coupled to metallic leads}
\author{A. Goker$^1$, B. A. Friedman$^2$ and P. Nordlander$^1$}

\address{$^1$
Department of Physics and Department of Electrical and Computer Engineering, \\
Rice Quantum Institute, Rice University, Houston, TX 77251-1892, USA
}
\address{$^2$
Department of Physics,
Sam Houston State University, Huntsville, TX 77341, USA
}
\date{\today}

\ead{nordland@rice.edu}

\begin{abstract}
The time-dependent non-crossing approximation is used to study 
the transient current in a single electron transistor attached 
asymmetrically to two leads following a sudden change in the
energy of the dot level. We show that for asymmetric coupling,
sharp features in the density of states of the leads can induce
oscillations in the current through the dot. These oscillations persist
to much longer timescales than the timescale for charge fluctuations.
The amplitude of the oscillations increases as the temperature or
source-drain bias across the dot is reduced and saturates for values
below the Kondo temperature.  We discuss the microscopic origin of these 
oscillations and comment on the possibility for their experimental detection.  
\end{abstract}

\pacs{72.15.Qm, 85.30.Vw, 73.50.Mx}

\maketitle

\section{Introduction}
Quantum effects are likely to play an increasing role in electronic devices as
the physical size of their components continues to shrink.
Quantum dots and qubits are examples of devices where quantum effects play a direct
role in their function. In a Single Electron Transistor (SET), i.e.,
a quantum dot coupled to two metallic leads, the conductance
can be drastically enhanced by the Kondo effect which can
occur when the quantum dot is populated by an odd number of
electrons \cite{NgLee88PRL,PustilnukGlazman04JPCM,GiulianoetAl04JPCM,%
GoldhaberetAl98Nature,CronenwettetAl98Science,HewsonetAl05JPCM,JiangSun07JPCM,%
GalperinetAl07JPCM}.
The Kondo effect is a quantum-coherent many-body state in which a spin 
singlet state is formed between the unpaired localized electron and 
delocalized electrons at the Fermi energy at low
temperatures \cite{Kondo64PTP}. 

An important issue for the function of any electronic device is how fast
it can respond to time-dependent perturbations and
bias \cite{ElzermanetAl04Nature,XingetAl07PRB}.
Several studies of time-dependent response of symmetrically-coupled SET
have been performed using a variety of methods
\cite{HettlerSchoeller95PRL,SchillerHershfield96PRL,%
GoldinAvishai98PRL,NordlanderetAl99PRL,KaminskyetAl99PRL,%
SchillerHershfield00PRB,KoganetAl04Science,LobaskinKehrein05PRB,%
MonrealFlores05PRB,GalperinetAl07PRB}.
In the case of a sudden switching of the dot level, the transient current
has been found to exhibit several distinct timescales
\cite{MerinoMarston04PRB,PlihaletAl05PRB,AndersSchiller05PRL,AndersetAl06PRB}. 
The fastest timescale is associated with charge relaxation and the other
much slower timescales are associated with the formation of a Kondo state,
i.e. spin dynamics. The detailed evolution of
the instantaneous currents following a sudden change of the dot level
has been shown to depend sensitively on external parameters such
as source-drain bias, external temperature, dot-lead coupling 
and position of the dot level. These above mentioned time-dependent
studies have all
been concerned with quantum dots which are symmetrically coupled to
their leads.
While the effect of asymmetric coupling on the steady-state conductance of an SET
has been studied both theoretically and
experimentally,\cite{KrawiecWysokinski02PRB,QuirionetAl06EPJB}
to our knowledge, the transient transport properties of an asymmetrically
coupled dot in the Kondo regime have not been investigated previously.

In this paper, we use a recently developed multi-scale
many-body transport method to study the effect
of asymmetric dot-lead coupling on the
transient transport in a quantum dot \cite{IzmaylovetAl06JPCM}.
We show that for a quantum dot asymmetrically coupled to two leads
with  sharp features in their Density of States (DOS),
the current can display sinusoidal modulations for timescales
well beyond the fast charge relaxation timescale.
The frequency of these sinusoidal modulations is given by the
energy difference between the Kondo resonance and the DOS feature.
The amplitude of the oscillations is found to increase
with decreasing temperature and source-drain bias and saturate for temperatures 
below the Kondo scale.
We attribute this phenomenon to an interference effect between 
the Kondo resonance at the Fermi level of the leads and the conduction
electrons around the DOS feature.
The magnitude of the oscillations depends sensitively on the
structure of the DOS feature of the leads. We show that
these oscillations also can occur for leads with a smooth DOS
but with finite bandwidth.

\section{Time dependent current in infinite-U Anderson model}

The SET is modeled as a single spin degenerate level of
energy $\varepsilon_{dot}$ coupled to leads through tunnel barriers,
\begin{eqnarray}
H(t)&=&\sum_{\sigma} \varepsilon_{dot}(t)n_{\sigma}
+\sum_{k\alpha\sigma}\varepsilon_{k}n_{k\alpha\sigma}
+{\textstyle\frac{1}{2}}{\sum} U n_{\sigma}n_{\sigma'}+ \nonumber \\
& &\sum_{k\alpha\sigma} \left[ V_{\alpha}(\varepsilon_{k\alpha})c_{k\alpha\sigma}^{\dag}c_{\sigma}+
{\rm h.c.} \right],
\label{Anderson}
\end{eqnarray}
where $c^\dagger_\sigma$ ($c_\sigma$) and $c^\dagger_{k\alpha\sigma}$ 
($c_{k\alpha\sigma}$) with $\alpha$=L,R create (annihilate) an electron 
of spin $\sigma$ in the dot level and in the left(L) and right(R) leads 
respectively. The $n_\sigma$ and $n_{k\alpha\sigma}$ are the corresponding
number operators and $V_{\alpha}$ are the hopping amplitudes for the left 
and right leads. The Coulomb repulsion energy $U$ is assumed to be sufficiently 
large that double occupancy of the dot level is prohibited. In the following, 
we will use atomic units with $\hbar=k_{\rm B}=e=1$.

To deal with large Coulomb correlation, we
use the auxiliary boson method for the Anderson Hamiltonian.
Briefly, the ordinary electron operators on the site are decomposed
into a massless boson operator and a pseudofermion operator as
\begin{equation}
c_{\sigma}=b^{\dagger} f_{\sigma},
\end{equation}
with the constraint that the number of massless "slave" bosons
plus the number of pseudofermions
\begin{equation}
Q=b^{\dagger}b+\sum_{\sigma}f^{\dagger}_{\sigma}f_{\sigma}
\end{equation}
must be equal to unity. Self-energies for the pseudofermion and massless
"slave" boson are then projected onto the physically relevant $Q=1$
subspace.

The Green's functions of the dot levels are calculated using the Non Crossing
Approximation (NCA) by numerical integration on a multi-scale time
grid \cite{IzmaylovetAl06JPCM}.  When the dot
level $\varepsilon_{dot}$ lies well below the Fermi level $\varepsilon_{F}$,
the spectral function of the dot (local density of states) exhibit two features:
a broad Fano-like resonance of full-width
\begin{equation}
\Gamma_{tot}(\varepsilon) =
2 \pi \sum_{k} (|V_{L} (\varepsilon_k)|^2 +|V_{R} (\varepsilon_k)|^2) 
\delta(\varepsilon-\varepsilon_{k})
\label{gamma} 
\end{equation}
around the dot level, and 
a sharp temperature sensitive 
resonance at the Fermi level (the Kondo peak), characterized by a low 
energy scale $T_K$ (the Kondo temperature)
\begin{equation}
T_K \propto \left(\frac{D\Gamma_{tot}}{4}\right)^\frac{1}{2}
\exp\left(-\frac{\pi|\varepsilon_{\rm dot}|}{\Gamma_{tot}}\right),
\label{tkondo}
\end{equation}
where $D$ is a high energy cutoff equal to half bandwidth
of the conduction electrons and $\Gamma_{tot}$ corresponds to
the value of $\Gamma_{tot}(\varepsilon)$ at $\varepsilon=\varepsilon_F$.
All energy units in this paper will be given in terms of $\Gamma_{tot}$. 

The current flowing through the SET can be calculated as the difference between
the currents from the left and right leads as
\begin{equation}
I(t)=I_L(t)-I_R(t),
\label{eq:ILR}
\end{equation}
The most general expression for the net current flowing from 
the left(right) lead is given by \cite{JauhoetAl94PRB}
\begin{eqnarray}
I_{L(R)}(t)&=&-2Im( \int_{-\infty}^{t} dt_1 \int \frac{d\varepsilon}{2\pi}
e^{-i\varepsilon(t_1-t)}\Gamma_{L(R)}(\varepsilon) \nonumber \\
& & e^{i\int_{t_1}^{t}d\tau \Delta_{L(R)}(\tau)}
[G^{<}(t,t_1)+f_{L(R)}(\varepsilon)G^{R}(t,t_1)])
\label{wingreen}
\end{eqnarray}
where the coupling functions $\Gamma_{L(R)}$
\begin{equation}
\Gamma_{L(R)}(\varepsilon)=2 \pi \rho_{L(R)}(\varepsilon)
V_{L(R)}(\varepsilon) V_{L(R)}^{*}(\varepsilon),
\end{equation}
depend on the DOS of the leads $\rho_{L(R)}(\varepsilon)$ and the hopping matrix
elements  in Eq.(\ref{Anderson}). The quantity
$\Delta_{L(R)}$ represents the time-dependence of the 
single particle energies in the left and right leads.
In this paper, we consider the case of a small constant
bias across the leads and no explicit time and energy dependence of the hopping
matrix elements, therefore $V_{L(R)}(\varepsilon)=V_{L(R)}(\varepsilon_f)$. 
We further make the assumption that the DOS of the leads
are the same, i.e. $\rho_{L(R)}(\varepsilon)=\rho(\varepsilon)$.
Thus the coupling functions can be parameterized
as,
\begin{equation}
\Gamma_{L(R)}(\varepsilon)=\bar{\Gamma}_{L(R)} \rho(\varepsilon)
\end{equation}
where $\bar{\Gamma}_{L(R)}$ are constants given by 
$\bar{\Gamma}_{L(R)}=2 \pi |V_{L(R)}(\varepsilon_f)|^2$ and
they determine the asymmetry of the couplings. In terms of
these constants, 
$\Gamma_{tot}=(\bar{\Gamma}_{L}+\bar{\Gamma}_{R})\rho(\varepsilon_f)$.

In Fig.~1 we show the functions $\rho(\varepsilon)$
that will be used to model the leads. 
The Lorentzians used in Fig.~1c and Fig.~1d to model sharp DOS features
have a width equal to 0.002~$\Gamma_{tot}$.
The DOS function $\rho(\varepsilon)$ has been normalized for all cases such
that the bands contain the same number of electrons.
The bandwidth of the leads is assumed to be 2$D$
with the Fermi energy at $\varepsilon=0$.

\begin{figure}[htb]
\centerline{\includegraphics[width=12cm]{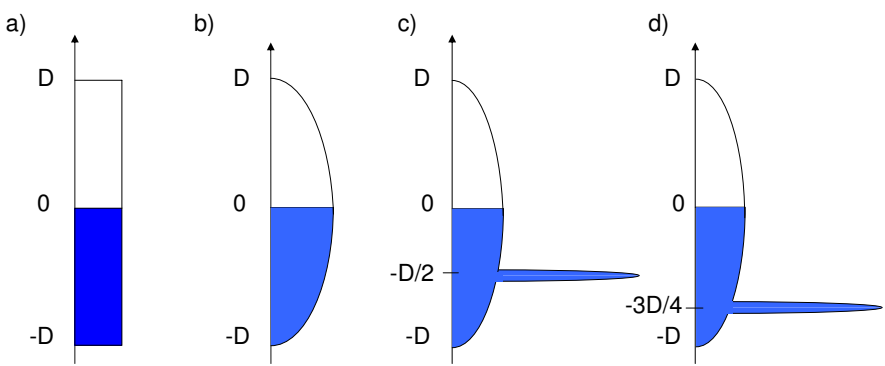}}
\caption{
This figure shows the functions $\rho(\varepsilon)$
used to model the leads.
Panel a) corresponds  to a rectangular band, b) to a parabolic band,
and c) and d) to a parabolic band with a Lorentzian feature
located at $\varepsilon$=-D/2
and $\varepsilon$=-3D/4 respectively
}
\end{figure}

The expression for the current can then be written as
\begin{eqnarray}
I(t)&=&-2\bar{\Gamma}_{L}
Im(\int_{-\infty}^{t} dt_1 (G^{<}(t,t_1)h(t-t_1)+G^{R}(t,t_1)f_{L}(t-t_1))) \nonumber \\
& &+2\bar{\Gamma}_{R}Im(\int_{-\infty}^{t} dt_1 
(G^{<}(t,t_1)h(t-t_1)+G^{R}(t,t_1)f_{R}(t-t_1))),
\label{current1}
\end{eqnarray}
where
\begin{equation}
h(t-t_1)=\int_{-D}^{D} \frac{d\varepsilon}{2\pi}\rho(\varepsilon)e^{i\varepsilon(t-t_1)},
\label{h}
\end{equation}
\begin{equation}
f_{L}(t-t_1)=\int_{-D}^{D} \frac{d\varepsilon}{2\pi}\rho(\varepsilon)
\frac{e^{i\varepsilon(t-t_1)}}{1+e^{\beta \left [\varepsilon-\frac{V}{2}\right]}},
\label{f}
\end{equation}
and
\begin{equation}
f_{R}(t-t_1)=\int_{-D}^{D} \frac{d\varepsilon}{2\pi}\rho(\varepsilon)
\frac{e^{i\varepsilon(t-t_1)}}{1+e^{\beta \left [\varepsilon+\frac{V}{2}\right]}}.
\label{fr}
\end{equation}
In these expressions, $V$ represents the source-drain bias.

The physical Green's functions in 
Eq.~(\ref{current1}) in can be expressed in terms of the pseudofermion and slave 
boson Green's functions using a projection approach discussed 
previously \cite{ShaoetAl194PRB}.
The final expression for the current is
\begin{eqnarray}
I(t)&=&-2\bar{\Gamma}_{L}\textit{Im} (\int_{-\infty}^{t} dt_1
(G_{pseudo}^{<}(t,t_1)B^{R}(t_1,t)h(t-t_1)\nonumber \\
& &-i((G_{pseudo}^{R}(t,t_1)B^{<}(t_1,t)+G_{pseudo}^{<}(t,t_1)
B^{R}(t_1,t))f_{L}(t-t_1)))\nonumber \\
& &+2\bar{\Gamma}_{R}\textit{Im} (\int_{-\infty}^{t} dt_1
(G_{pseudo}^{<}(t,t_1)B^{R}(t_1,t)h(t-t_1)\nonumber \\
& &-i((G_{pseudo}^{R}(t,t_1)B^{<}(t_1,t)+G_{pseudo}^{<}(t,t_1)
B^{R}(t_1,t))f_{R}(t-t_1))).
\label{final} 
\end{eqnarray}
This is the main result of this section and will be used
in the calculations presented below.
  
\section{Results}

In this section we will analyze the instantaneous current
following a sudden change of the dot level from a position
at $\varepsilon_1$=5$\Gamma_{tot}$ below the Fermi
level where, for the present finite temperatures, the Kondo effect will
be absent to a position at $\varepsilon_2$=2$\Gamma_{tot}$
closer to the Fermi energy where the Kondo effect will be present.
For a parabolic DOS function (Fig.~1b) with $D$=9$\Gamma_{tot}$,
the Kondo temperature in the final state is approximately
$T_K=$0.0016$\Gamma_{tot}$.

\begin{figure}[htb]
\centerline{\includegraphics[width=12cm]{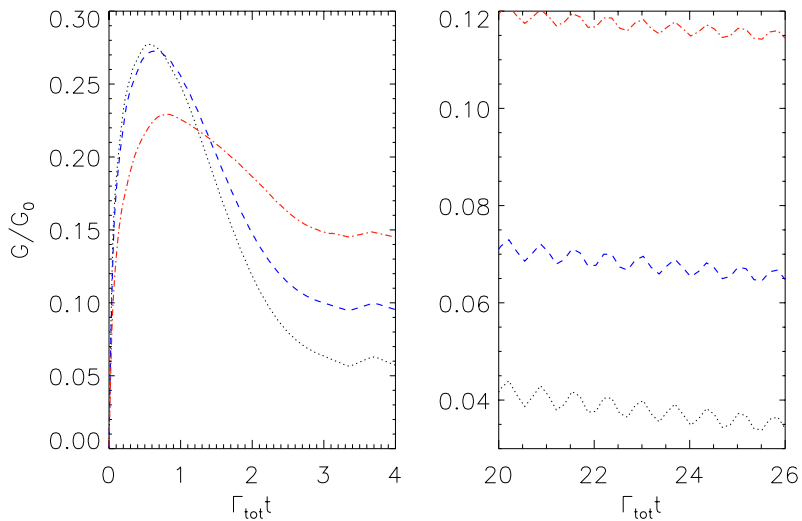}}
\caption{
Black(dotted), blue(dashed) and red(dot dashed) curves in panel a show the
instantaneous conductance vs. time for rectangular DOS
in Fig. 1a immediately after the dot level is switched to its final position
for asymmetry factors of 0.95,0.9 and 0.85 respectively.
$\Gamma_{tot}$ is fixed with D=9$\Gamma_{tot}$ at T=0.0093$\Gamma_{tot}$
and V=$T_K$. The beginning of the oscillations is clearly visible in this panel.
Black(dotted), blue(dashed) and red(dot dashed) curves in panel b are the continuation
of the ones in the first panel in the long timescale for the same parameters.
}
\label{asymmetry}
\end{figure}

We begin our analysis by investigating the effect of 
asymmetric coupling to the leads. We define the asymmetry factor as the
the ratio $\bar{\Gamma}_{L}/\bar{\Gamma}_{tot}$ 
where $\bar{\Gamma}_{tot}$= $\bar{\Gamma}_{L}+\bar{\Gamma}_{R}$.
In Fig.~2, the instantaneous conductance, $G=I(t)/V$, is plotted as a function of time
after the dot level is switched for various asymmetry factors with small bias.
The final steady state conductances are in perfect agreement with
previous theoretical results \cite{WingreenMeir94PRB}.
The transient short timescale associated with $\Gamma_{tot}$ shown
in Fig.~2a is due to the formation of the broad Fano-like
resonance at $\varepsilon_2$. The transient increase in the instantaneous
current is due to the charging of the dot. The steady state current is determined by the asymmetry
factor. For an asymmetry factor of 1 where the steady state current is zero, we have verified the 
accuracy of our numerical approach by showing that the integrated current is equal to the change of
the charge of the dot level.

Fig.~2b shows the instantaneous conductances
for larger times on a magnified scale. It is clear from this panel that the
current exhibits sinusoidal modulations at timescales well beyond
$\Gamma_L$ or $\Gamma_R$.  As we reduce the 
asymmetry factor, the amplitude of these sinusoidal modulations 
starts decreasing and eventually disappears for symmetric coupling.
The frequency of the conductance oscillations is
equal to the bandwidth $D$ of the leads. External parameters such as 
the energy or width of the dot level, asymmetry factor, and ambient temperature
and source-drain bias only influence the amplitude of the oscillations.

\begin{figure}[htb]
\centerline{\includegraphics[width=12cm]{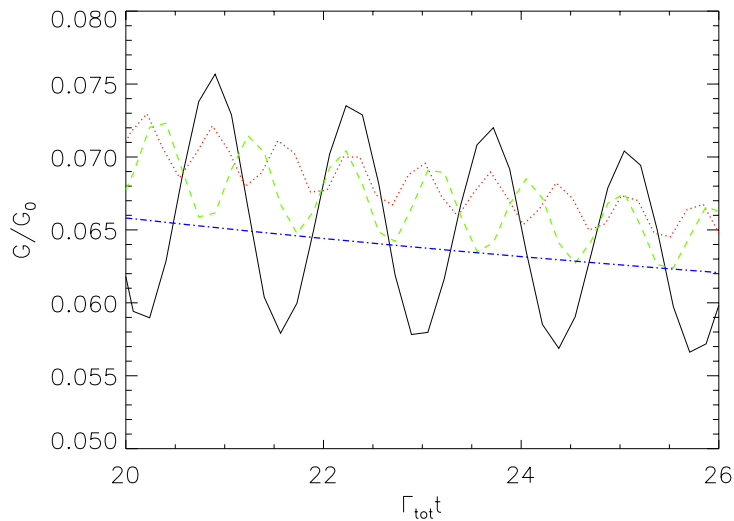}}
\caption{
Effect of the DOS function on the instantaneous conductance.
Red(dotted), dark blue(dot dashed), black(solid) and green(dashed) curves display
the instantaneous conductance vs. time in the long timescale for cases a, b, c and d
respectively in Fig.1 when the source-drain bias is equal to V=$T_K$
with fixed $\Gamma_{tot}$. All curves are for asymmetry factor of 0.9,
D=9$\Gamma_{tot}$ at T=0.0093$\Gamma_{tot}$.
}
\label{dos}
\end{figure}

In Fig.~3 we show the effect of the the DOS function of the leads
$\rho(\epsilon)$ on the conductance oscillations. 
The oscillations do not look like perfectly sinusoidal functions
because of our use of a finite time step in the numerical solution
of the Dyson equations.  However, the results are fully converged
and for a finer time steps we recover almost perfectly sinusoidal oscillations.
Figure~3 reveals that the 
DOS can have a pronounced effect on the conductance oscillations.
The largest oscillations occur for a DOS with a peak feature as
in Fig.~1c and 1d. For a parabolic DOS function (Fig.~1b),
conductance oscillations are still present but not discernible on the scale
of the figure. The frequency of the current oscillations is equal to
the energy difference between the Fermi level of the leads and the feature
in the DOS function. For the rectangular and parabolic DOS
function (Fig.~1a and 1b) where the DOS feature is the band cut-off,
the period is equal to 1/$D$=1/9 and for the DOS functions
depicted in Fig.~1c and 1d where the DOS feature is the narrow
Lorentzian peak, the periods are 2/$D$ and 4/3$D$ respectively.

We have also investigated other shapes of DOS functions.
For instance, for a triangular DOS,  $\rho(\varepsilon)=2(1+\varepsilon/D)$, the conductance
displays oscillations whose frequency is equal to $D$ very similar to
what was obtained for the parabolic DOS in Fig.~1b.
For two sharp features at different energies $D_1$ and $D_2$ in the DOS 
we obtain conductance modulations of frequencies proportional
to $(D_1+D_2)$ and $(D_1-D_2)$.

Figure.~4 shows the  effect of temperature and source-drain 
bias on the conductance oscillations. The figure demonstrates that
the amplitude of the oscillations decrease with temperature and
bias across the dot but their frequency remain unchanged.
The figure shows two important effects. First, it takes longer for the current
and thus the damped oscillations to decay as the temperature or
source-drain bias is reduced.  Second, the amplitude 
of the damped oscillations increases with decreasing temperature
or source-drain bias and but saturates for temperatures and bias below
the Kondo temperature, which we estimate to be 
around $T_K=$0.0016$\Gamma_{tot}$ in these systems.

\begin{figure}[htb]
\centerline{\includegraphics[width=12cm]{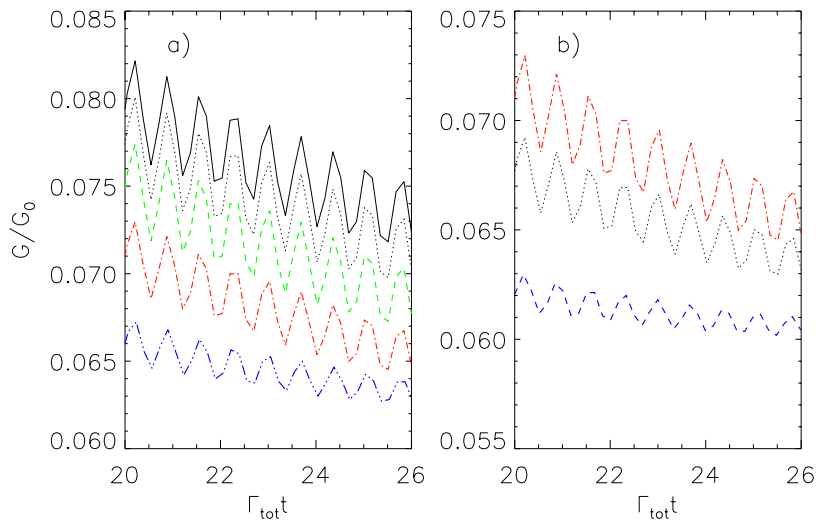}}
\caption{
Panel a shows the instantaneous conductance for
several different temperatures with asymmetry factor of 0.9,
fixed $\Gamma_{tot}$, D=9$\Gamma_{tot}$ and rectangular
DOS. Dark blue(dot dot dashed), red(dot dashed), green(dashed),
light blue(dotted) and black(solid) curves represent the conductance at T=0.0186,
0.0093, 0.0046, 0.0023, and 0.0009$\Gamma_{tot}$ respectively.
Red(dot dashed), black(dotted) and dark blue(dashed) curves in panel b display
the instantaneous conductance vs. time in the long timescale for rectangular
DOS when the source-drain bias is equal to V=$T_K$, V=5$T_K$ and
V=10$T_K$ respectively with fixed $\Gamma_{tot}$ and D=9$\Gamma_{tot}$
for asymmetry factor of 0.9 at T=0.0093$\Gamma_{tot}$.
}
\label{bias}
\end{figure}

\section{Discussion}

Our numerical calculations clearly show that the timescale for the decay of the
conductance oscillations is much larger than the fast timescales set by the
couplings of the dot level to the leads, $\Gamma_L$ or $\Gamma_L$.
The timescale does not depend on the width or shape of the DOS feature in the
leads.  It appears that the timescale is related to the Kondo resonance.
When the Kondo resonance is fully formed at times around $1/T_K$, the oscillations
disappears.
 
Further support for the role of Kondo physics in these conductance oscillations
comes from the observation of a
saturation of their amplitude for temperatures
and source-drain bias below the Kondo temperature (Fig.~4).
The effect of temperature and source-drain bias on the Kondo resonance in an SET
was recently investigated in detail \cite{PlihaletAl05PRB}. For values smaller than
the Kondo temperature, the Kondo resonance is fully formed just above the Fermi Level.
Increasing the temperature above $T_K$ broadens and reduces the magnitude of the
resonance. Increasing the source-drain bias results in the formation of a split Kondo
resonance with strongly reduced intensities. For temperatures or source-drain
bias well above the Kondo temperature, the Kondo resonance is completely
suppressed. Since the frequency of the
oscillations is determined by the energy difference between the Fermi level and
the DOS feature, we believe that the oscillations reflect an 
interference process between the conduction electrons associated with the Kondo
resonance and the conduction electrons associated with the DOS feature.
In order to substantiate this explanation we used a simple analytical
noninteracting Anderson model to calculate the instantaneous
conductance following the sudden switching of a dot level of a width $\Gamma_{eff}$
equal to that of the final Kondo resonance in the interacting model.
This effective resonance was switched to a position just above the Fermi level
of the leads. These calculations were performed numerically
for both the parabolic and rectangular DOS functions shown in Fig.~1a and 1b.
The instantaneous 
conductance was found to display the same conductance oscillations with frequencies 
determined by the bandwidth $D$ as in the original interacting model. The oscillations 
are found to decay over a long timescale determined by $\Gamma_{eff}$.

While the Kondo temperature depends only on the total coupling to the
leads $\Gamma_{tot}$, the total current depends on the asymmetry factor.
The reason why the conductance oscillations only show up for asymmetric
couplings is that the total current through the dot
is the difference of the currents from the left and right leads Eq.~(\ref{eq:ILR}).
The interference effect shows up both in the the right and left currents.
For symmetric coupling, the current oscillations from the right and left leads
are out of phase resulting in a cancellation of the oscillations
in the total current. For asymmetric coupling the left and right current oscillations
do not cancel resulting in the observed conductance oscillations.
This result has also been verified numerically for the noninteracting model
described in the preceding paragraph.

For dots coupled to leads with very weak DOS features,
the conductance oscillations will be very small. In Fig.~5 we show the
time-derivative of the instantaneous current for leads with a
parabolic DOS function (Fig.~1b) for three different values
of the bandwidth $D$.  The conductance oscillations here
results from the interference of the Kondo resonance in the final state
and the weak discontinuity in the DOS at the lower band edge
of the leads.  The figure clearly demonstrates oscillations of the instantaneous
currents with a frequency equal to $D$ in the long timescale.

\begin{figure}[htb]
\centerline{\includegraphics[width=12cm]{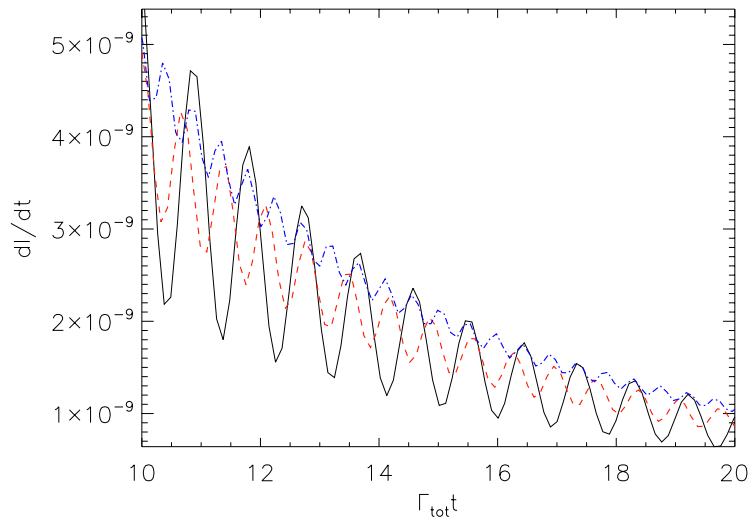}}
\caption{
Time-derivative of the instantaneous current in the long timescale
for parabolic DOS (Fig.~1b). Black(solid), red(dashed)
and dark blue(dot dashed) curves correspond to D=6.75$\Gamma_{tot}$,
D=9$\Gamma_{tot}$ and D=13.5$\Gamma_{tot}$ for asymmetry
factor of 1.0 at T=0.0093$\Gamma_{tot}$.
}
\end{figure}

The experimental study of the conductance oscillations could possibly
be made using previously suggested techniques, i.e., by measuring the total 
charge transport as function of pulse duration \cite{NordlanderetAl99PRL}.
For a bandwidth of $D$=1~eV, the oscillation period will be of
an order of $10^{-14}$ seconds. Since the current oscillates, electromagnetic
radiation will be emitted at a rate proportional to $[\frac{dI(t)}{dt}]^2$.
For a suitably designed system it may be possible to detect the emitted light.
For $D$=1~eV the emission would occur in the 
infrared at a photon energy of 0.1~eV.

\section{Conclusion}
In this paper, we employed the time dependent non-crossing approximation
to analyze the transient current in a single electron transistor in the
Kondo regime asymmetrically coupled to two metallic leads with features
in their DOS.
We show that for asymmetric coupling, the conductance can exhibit oscillations
which persist for times much longer than the timescale for charge relaxation.
The origin of these oscillations is an interference between the conduction
electrons associated with the Kondo resonance and those associated with the
DOS feature. The amplitude of the oscillations are found to depend strongly on
temperature and source-drain bias  when those exceed the Kondo temperature.
We hope that our 
predictions will motivate further theoretical and experimental studies.

\section*{Acknowledgments}
We thank Prof. Hong Guo for communication 
regarding relevant work \cite{MaciejkoetAl06PRB}.  
This work was supported by the Welch Foundation under grant C-1222.

\clearpage

\providecommand{\newblock}{}

\end{document}